\documentclass{article}

\usepackage[final, dblblindworkshop]{neurips_2025}

\usepackage[utf8]{inputenc} % allow utf-8 input
\usepackage[T1]{fontenc}    % use 8-bit T1 fonts
\usepackage{hyperref}       % hyperlinks
\usepackage{url}            % simple URL typesetting
\usepackage{booktabs}       % professional-quality tables
\usepackage{amsfonts}       % blackboard math symbols
\usepackage{nicefrac}       % compact symbols for 1/2, etc.
\usepackage{microtype}      % microtypography
\usepackage[table]{xcolor}  % colors, with table support
\usepackage{amsmath}        % for math environments
\usepackage{graphicx}
\usepackage{natbib}         % for citations
\usepackage{multirow}       % for multi-row tables
\usepackage{siunitx}        % for number formatting
\usepackage{amssymb}        % for math symbols
\usepackage{enumitem}
\usepackage{adjustbox}      % for resizing tables
\usepackage{float}

\definecolor{mygreen}{RGB}{0, 150, 0}
\definecolor{myred}{RGB}{200, 0, 0}
\newcommand{\gcolor}[1]{{\color{mygreen}#1}}
\newcommand{\rcolor}[1]{{\color{myred}#1}}

\workshoptitle{AI for Tabular Data workshop at EurIPS 2025}
\title{Exploring Multi-Table Retrieval Through Iterative Search}

\author{%
  Allaa Boutaleb, Bernd Amann, Rafael Angarita and Hubert Naacke \\
  \normalsize Sorbonne Université, CNRS, LIP6, F-75005 Paris, France \\
  \texttt{\{firstname.lastname\}@lip6.fr}
}

\begin{document}

\maketitle
\begin{abstract}
Open-domain question answering over datalakes requires retrieving and composing information from multiple tables, a challenging subtask that demands semantic relevance and structural coherence (e.g., joinability). While exact optimization methods like Mixed-Integer Programming (MIP) can ensure coherence, their computational complexity is often prohibitive. Conversely, simpler greedy heuristics that optimize for query coverage alone often fail to find these coherent, joinable sets. This paper frames multi-table retrieval as an iterative search process, arguing this approach offers advantages in scalability, interpretability, and flexibility. We propose a general framework and a concrete instantiation: a fast, effective Greedy Join-Aware Retrieval algorithm that holistically balances relevance, coverage, and joinability. Experiments across 5 NL2SQL benchmarks demonstrate that our iterative method achieves competitive retrieval performance compared to the MIP-based approach while being 4-400x faster depending on the benchmark and search space settings. This work highlights the potential of iterative heuristics for practical, scalable, and composition-aware retrieval.
\end{abstract}

\section{Introduction}

Large Language Models (LLMs) increasingly leverage Retrieval-Augmented Generation (RAG) pipelines for natural language interfaces to structured databases~\citep{li-etal-2023-bird, wang2024redefining}. In this paradigm, the accuracy and coherence of an LLM's answer is critically dependent on the retrieval phase: if the retrieved tables do not collectively contain the necessary information or cannot be coherently composed (e.g., via joins), the LLM's ability to generate a correct answer is severely compromised.
While previous work on table retrieval has primarily focused on retrieving individual tables from large corpora~\citep{herzig-etal-2021-open}, complex real-world queries often require retrieving and composing information from multiple tables—a challenge known as \textbf{multi-table retrieval}. This task introduces significant challenges beyond simple relevance ranking. Consider a query like: \textit{"Show names and total order values for New York customers who ordered 'Laptop' and any customers who ordered 'Smartphone' after 2024."} Answering this requires tables like Customers, Orders, and Products, but the retrieval process faces several issues.
Due to data redundancy or different valid reasoning paths, \textbf{multiple valid sets} of tables might answer the query. Furthermore, \textbf{path dependency} can arise in iterative methods, where the order of table selection influences the final set.
This inherent ambiguity suggests that multi-table retrieval is not always about finding a single, pre-defined optimal set, but often involves exploring multiple plausible evidence paths.

Traditional retrieval methods, which assess tables independently, struggle with structural composability. \citet{chen-etal-2024-table} addressed this by proposing \textbf{composition-aware} retrieval focused on joinability, reframing the task as selecting a coherent sub-graph. Their Join-Aware Retrieval (JAR) method performs neural retrieval followed by a Mixed-Integer Program (MIP) to find the provably optimal joinable subset. Although highly effective, this one-shot NP-hard optimization suffers in terms of scalability, as acknowledged by the authors. JAR has inspired follow-ups combining solver logic with LLMs~\citep{chen-etal-2025-arm} and agentic, multi-hop reasoning approaches like MURRE~\citep{zhang-etal-2025-murre}. Other related works like CRUSH4SQL~\citep{kothyari-etal-2023-crush4sql} use a greedy decomposition strategy, but lack the explicit join-aware component central to JAR. Recent works such as DBCopilot~\citep{DBLP:conf/edbt/WangCLH0WZ25} also explore iterative retrieval via schema routing for massive databases.
Complementary to these efforts, recent work has also proposed retrieval-augmented methods that enable LLMs to interact with structured data via iterative and adaptive search frameworks~\citep{wang2024redefining}.

This paper proposes framing multi-table retrieval as an \textbf{iterative and explorative search process}. To stimulate discussion on practical, scalable architectures for this complex retrieval task, we posit that constructing the table set step-by-step offers benefits in scalability (via heuristics), interpretability, and flexibility.
Our contributions are threefold: (1) We propose a general, flexible iterative framework for multi-table retrieval. (2) We detail a fast and effective \textbf{Greedy Join-Aware Retrieval} algorithm as a concrete instantiation of this framework. (3) We demonstrate its empirical viability on standard benchmarks (\citet{yu-etal-2018-spider}, \citet{li-etal-2023-bird}) as well as more complex enterprise benchmarks (\citet{10.14778/3407790.3407858}, \citet{chen2024beaver}), showing it achieves competitive performance to the MIP-based JAR approach while being over \num{4}-\num{400}x faster\footnote{Code available at: \url{https://github.com/Allaa-boutaleb/iterative-jar/}}.

\section{An Iterative Framework for Multi-Table Search}

We propose framing multi-table retrieval as a sequential decision-making process where an algorithm iteratively expands a set of selected tables based on a dynamic \textbf{context}. This approach offers several advantages over one-shot global optimization:
\begin{itemize}[nosep]
    \item \textbf{Interpretability:} Each selection step provides a checkpoint for analyzing the reasoning process, potentially enabling human-in-the-loop guidance.
    \item \textbf{Extensibility:} The framework is modular and the selection logic can be dynamically adapted to prioritize different operators (e.g., JOIN vs. UNION) or evolving objectives.
    \item \textbf{Heuristic Potential:} While global optimization is often NP-hard for this task, step-by-step construction lends itself to efficient polynomial-time heuristics, addressing scalability.
\end{itemize}

\paragraph{Representing Query Coverage} A central challenge in multi-table retrieval is tracking the capacity to answer the initial query by the already selected tables. To illustrate, consider the query \textit{"For movies with the keyword of 'civil war', calculate the average revenue generated by these movies"}. To estimate what information is needed, methods process this query using an LLM in different ways. For instance, \citet{kothyari-etal-2023-crush4sql} use the LLM to "hallucinate" a minimal, potential schema (e.g., \texttt{movies(title, revenue)}) that could answer it. In contrast, our method (following \citet{chen-etal-2024-table}) decomposes the query into a set of fine-grained concepts or sub-queries $\{q_j\}$, such as \{\texttt{movies:keyword}, \texttt{movies:revenue}\}. This general idea of estimating the query's requirements is known as \emph{query coverage}. In our framework, query coverage tracks how well each concept $q_j$ is addressed by the tables selected so far.

\paragraph{Abstract Formulation}
Let $\mathcal{T}$ be the set of all available candidate tables. The search process evolves at each step $k$ a \textbf{context}
$\mathcal{C}_k = (G_k, \mathbf{Q}_k)$ where:
\begin{itemize}
    \item $G_k = (S_k, E_k)$ is a graph of $k$ selected tables $S_k \subseteq \mathcal{T}$ and their discovered relationships $E_k$ (e.g., potential joins).
    \item $\mathbf{Q}_k$ represents the query coverage state for the query concepts $\{q_j\}$. For example, in a simple setting we can represent $\mathbf{Q}_k$ by a vector $\mathbf{q}_k$ whose size equals the number of concepts, where element $q_{k,j}$ quantifies coverage for concept $q_j$.
\end{itemize}
The process starts with the empty context $\mathcal{C}_0 = ((\emptyset, \emptyset), \mathbf{0})$ with no selected tables and no coverage.
At each step $k$, a \textbf{selection function} $\Phi$ chooses the next table $T_{k+1}$ from the remaining candidates $\mathcal{T} \setminus S_k$ by maximizing a \emph{context-dependent utility function} $U$: $T_{k+1} = \Phi(\mathcal{C}_k, \mathcal{T} \setminus S_k) = \arg\max_{T_i \in \mathcal{T} \setminus S_k} U(T_i, \mathcal{C}_k)$.
An \textbf{update function} $\Psi$ then transitions the system to the next state $\mathcal{C}_{k+1}$ by incorporating $T_{k+1}$: $\mathcal{C}_{k+1} = \Psi(\mathcal{C}_k, T_{k+1}) = (G_{k+1}, \mathbf{Q}_{k+1})$.
This process repeats until a stopping criterion is met (e.g., $k=K$ or $\min_j q_{k,j} \ge \theta$). We illustrate a concrete instantiation of the selection and update functions in the next section.

\section{Case Study: Greedy Join-Aware Retrieval}
In this section we present a simple instantiation of the iterative framework focused on \textit{join-aware} multi-table retrieval.
%\subsection{Pre-computed Input Signals}
Our algorithm, like JAR \citep{chen-etal-2024-table}, operates on pre-computed scores quantifying relevance and compatibility for a query $Q$ and candidate tables $\mathcal{T}$. These include: 
\begin{itemize}[nosep]
    \item \textbf{Coarse-grained Relevance ($r_i$)}, the overall $Q$-to-table $T_i$ semantic similarity via dense retriever embeddings \citep{izacard-etal-2022-contriever}, where $\text{emb}(\cdot)$ is the embedding function:
\begin{equation}
r_i = \cos(\text{emb}(Q), \text{emb}(T_i)) \in [-1, 1]
\end{equation}

\item \textbf{Fine-grained Relevance ($F_{ji}$)}, which measures how well $T_i$ addresses a specific sub-query $q_j$ (decomposed from $Q$ via an LLM) by finding the maximum similarity between $q_j$ and any column $c$ in $T_i$: $F_{ji} = \max_{c \in \text{cols}(T_i)} \cos(\text{emb}(q_j), \text{emb}(c)) \in [-1, 1]$.
\item \textbf{Join Compatibility ($\omega_{il}$)}, a score in $[0, 1]$ quantifying the join likelihood between $T_i$ and $T_l$ by combining schema, value overlap, and uniqueness signals to approximate a PK-FK link, following \citet{chen-etal-2024-table}.
\end{itemize}
%\subsection{The Algorithm}
The \textbf{context} is  $\mathcal{C}_k = (G_k, \mathbf{Q}_k)$ where $G_k = (S_k, E_k)$ is the graph of selected tables $S_k$ and their join paths $E_k$. The \textbf{coverage state} $\mathbf{Q}_k$ is implemented as the vector $\mathbf{q}_k$ storing the maximum $F_{ji}$ score seen for each sub-query $q_j$.
The algorithm's \textbf{selection function} $\Phi$ works by maximizing a utility function $U$ that is a weighted sum of three marginal gain components. These gains are calculated for a candidate table $T_i$ relative to the prior context $\mathcal{C}_{k-1} = (G_{k-1}, \mathbf{Q}_{k-1})$: 
\begin{itemize}
\item  \textbf{Coarse Relevance Gain}, $G_{\text{coarse}}(T_i) = r_i$, which is intrinsic to the table;
\item  \textbf{Marginal Coverage Gain}, which depends on the prior coverage vector $\mathbf{q}_{k-1}$:
\begin{equation}
G_{\text{cov}}(T_i | \mathcal{C}_{k-1}) = \sum_{j} \max(0, F_{ji} - (\mathbf{q}_{k-1})_j)
\end{equation}

\item  \textbf{Marginal Join Gain}, $G_{\text{join}}(T_i | \mathcal{C}_{k-1}) = \sum_{T_l \in S_{k-1}} \omega_{il}$, which depends on the nodes $S_{k-1}$ of the prior graph $G_{k-1}$. For the main iterative step ($k > 1$), the utility function $U(T_i, \mathcal{C}_{k-1})$ combines these gains:
\begin{equation}
T_k = \arg\max_{T_i \in \mathcal{T} \setminus S_{k-1}} \left[ \lambda_{\text{coarse}} G_{\text{coarse}}(T_i) + \lambda_{\text{cov}} G_{\text{cov}}(T_i | \mathcal{C}_{k-1}) + \lambda_{\text{join}} G_{\text{join}}(T_i | \mathcal{C}_{k-1}) \right]
\end{equation}
\end{itemize}

The seed selection ($k=1$) is a special case. Starting from $\mathcal{C}_0 = ((\emptyset, \emptyset), \mathbf{0})$, the $G_{\text{join}}$ term is undefined and $G_{\text{cov}}$ simplifies (as $\mathbf{q}_0 = \mathbf{0}$), so the utility function reduces to selecting based on individual merit:
\begin{equation}
T_1 = \arg\max_{T_i \in \mathcal{T}} \left[ \lambda_{\text{coarse}} \cdot r_i + \lambda_{\text{cov}} \cdot \sum_{j} F_{ji} \right]
\end{equation}
Finally, the \textbf{update function} $\Psi$ transitions the state to $\mathcal{C}_k$ by updating the graph $G_k = (S_{k-1} \cup \{T_k\}, E_{k-1} \cup E_{\text{new}})$ and the coverage vector $\mathbf{q}_k = \max(\mathbf{q}_{k-1}, \mathbf{F}_k)$, where $\mathbf{F}_k$ is the vector of fine-grained scores for $T_k$.

\section{Experiments and Analysis}
\paragraph{Experimental Setup.}
We compare our iterative greedy algorithm against three baselines: a \textbf{Dense Retrieval (Contriever)} baseline \citep{izacard-etal-2022-contriever}, the official \textbf{JAR$_{MIP}$} implementation\footnote{\url{https://github.com/peterbaile/jar}} \citep{chen-etal-2024-table}, and \textbf{CRUSH}~\citep{kothyari-etal-2023-crush4sql} (using Contriever as the embedding model).
For all methods, we use a consistent table schema serialization. The Contriever baseline embeds this serialized schema, the query, and its concepts within the same embedding space to retrieve an initial ranked list of tables.
For JAR (MIP) and our iterative method, we re-rank this initial candidate set: the \textbf{top-20} tables for SPIDER, BIRD, and BEAVER, and the \textbf{top-30} for FIBEN. Both methods use the identical set of pre-computed relevance ($r_i, F_{ji}$) and join compatibility ($\omega_{il}$) scores, ensuring a fair comparison of the selection algorithms.
For CRUSH, we follow the default settings from its implementation\footnote{ \url{https://github.com/tshu-w/DBCopilot/blob/master/scripts/crush4sql.py}}, using an initial candidate cutoff of 100 tables and a selection budget of 20. 
%  The lower performance of CRUSH (see Table \ref{tab:retrieval-performance}) is attributable to two core design differences: the algorithm is not explicitly join-aware, and it employs a different, more aggressive greedy selection logic focused on segment coverage, which can be sensitive to noise (see Appendix \ref{sec:crush_failure} for a detailed example).

We evaluate on the multi-table queries from SPIDER \citep{yu-etal-2018-spider}, BIRD \citep{li-etal-2023-bird}, FIBEN \citep{10.14778/3407790.3407858}, and the complex BEAVER \citep{chen2024beaver} benchmark (see Table \ref{tab:retrieval-performance} for statistics). All benchmarks are used in an \textbf{open-domain setting}; for SPIDER and BIRD, we use the specific versions provided by \citet{chen-etal-2024-table} which contain only the databases used in the test queries. BEAVER is composed of two distinct enterprise data warehouses: BEAVER-DW (university physical administration) and BEAVER-NW (virtual machine and network infrastructure). As our study focuses on multi-table retrieval, we exclude all single-table queries from our evaluation. A detailed breakdown of the multi-table query distribution for each benchmark is available in Appendix \ref{sec:distribution_of_queries}.
For our greedy method, we use weights $\lambda_{\text{cov}}=2.0, \lambda_{\text{join}}=1.0, \lambda_{\text{coarse}}=4.0$. These weights were determined empirically via an ablation study (Appendix \ref{sec:ablation_appendix}), which indicated that coarse relevance is the strongest signal, followed by coverage and join compatibility.
We report \textbf{Recall (R)} and \textbf{Complete Recall (CR)} for K $\in \{2, 3, 5, 10\}$, where CR is a binary set-level metric indicating if all ground-truth tables were retrieved. We use a \textbf{1 hour} per-query timeout for the MIP solver to allow it ample time to find optimal solutions.

\begin{table}[htbp]
  \caption{Retrieval performance on multi-table queries. K is the number of tables retrieved. Contriever shows absolute R/CR scores (\%) and base retrieval time (s). All other methods show relative R/CR gain (\gcolor{+}) or loss (\rcolor{-}) and total re-ranking time (+ s). \textdagger~indicates prohibitive runtime. \textbf{Highest} and \underline{Second-Highest} scores are marked per column within each benchmark.}
  \label{tab:retrieval-performance}
  \centering
  \begin{adjustbox}{width=\textwidth}
    \scriptsize 
    \setlength{\tabcolsep}{4pt} 
    \begin{tabular}{ll rr rr rr rr r}
      \toprule
      \multirow{2}{*}{\textbf{Benchmark}} & \multirow{2}{*}{\textbf{Method}} & \multicolumn{2}{c}{\textbf{K=2}} & \multicolumn{2}{c}{\textbf{K=3}} & \multicolumn{2}{c}{\textbf{K=5}} & \multicolumn{2}{c}{\textbf{K=10}} & \multirow{2}{*}{\textbf{Time (seconds)}} \\
      \cmidrule(lr){3-4} \cmidrule(lr){5-6} \cmidrule(lr){7-8} \cmidrule(lr){9-10}
      & & Recall (R) & CR & Recall (R) & CR & Recall (R) & CR & Recall (R) & CR & \\
      \midrule
      
      \multirow{4}{*}{\parbox{2.8cm}{\raggedright \textbf{SPIDER} \\[3pt] \tiny \linespread{0.9}\selectfont
        \texttt{Num. of DBs: 20} \\
        \texttt{Num. of Tables: 81} \\
        \texttt{Avg. Table Width: 5.4} \\
        \texttt{Num. of Queries: 459}}} 
      & Contriever & 81.3 & 59.9 & 93.8 & 85.6 & \underline{98.9} & \underline{97.6} & \textbf{99.7} & \textbf{99.3} & 15 \\
      \cmidrule(l){2-11}
      & JAR (MIP) & \gcolor{\underline{+4.1}} & \gcolor{\underline{+7.2}} & \gcolor{\textbf{+2.6}} & \gcolor{\textbf{+5.7}} & \gcolor{\textbf{+0.5}} & \gcolor{\textbf{+1.1}} & \textdagger & \textdagger & + (325 / 290 / 6360 / \textdagger) \\
      & \textbf{JAR}$_{\text{iterative}}$ \textbf{(Ours)} & \gcolor{\textbf{+4.2}} & \gcolor{\textbf{+8.1}} & \gcolor{\underline{+1.8}} & \gcolor{\underline{+4.4}} & \rcolor{-0.9} & \rcolor{-2.0} & \rcolor{\underline{-1.4}} & \rcolor{\underline{-2.8}} & + \textasciitilde20 \\
      & CRUSH$_{\text{Contriever}}$ & \rcolor{-17.6} & \rcolor{-25.3} & \rcolor{-21.4} & \rcolor{-38.1} & \rcolor{-21.5} & \rcolor{-42.0} & \rcolor{-21.9} & \rcolor{-43.3} & + \textasciitilde38 \\
      \midrule
      
      \multirow{4}{*}{\parbox{2.8cm}{\raggedright \textbf{BIRD} \\[3pt] \tiny \linespread{0.9}\selectfont
        \texttt{Num. of DBs: 11} \\
        \texttt{Num. of Tables: 75} \\
        \texttt{Avg. Table Width: 10.6} \\
        \texttt{Num. of Queries: 1172}}} & Contriever & 63.5 & 34.8 & 76.2 & 56.2 & 85.2 & 71.3 & \underline{95.1} & \underline{89.9} & 23 \\
      \cmidrule(l){2-11}
      & JAR (MIP) & \gcolor{\textbf{+10.9}} & \gcolor{\textbf{+16.9}} & \gcolor{\textbf{+9.4}} & \gcolor{\textbf{+15.7}} & \gcolor{\underline{+5.2}} & \gcolor{\underline{+10.4}} & \textdagger & \textdagger & + (8501 / 13679 / 40756 / \textdagger) \\
      & \textbf{JAR}$_{\text{iterative}}$ \textbf{(Ours)} & \gcolor{\underline{+10.7}} & \gcolor{\underline{+16.7}} & \gcolor{\underline{+8.5}} & \gcolor{\underline{+14.3}} & \gcolor{\textbf{+5.8}} & \gcolor{\textbf{+10.8}} & \gcolor{\textbf{+1.1}} & \gcolor{\textbf{+2.7}} & + \textasciitilde101 \\
      & CRUSH$_{\text{Contriever}}$ & \rcolor{-12.8} & \rcolor{-12.8} & \rcolor{-13.5} & \rcolor{-17.3} & \rcolor{-12.4} & \rcolor{-16.9} & \rcolor{-20.6} & \rcolor{-32.8} & + \textasciitilde116 \\
      \midrule

      \multirow{3}{*}{\parbox{2.8cm}{\raggedright \textbf{FIBEN} \\[3pt] \tiny \linespread{0.9}\selectfont
        \texttt{Num. of DBs: 1} \\
        \texttt{Num. of Tables: 152} \\
        \texttt{Avg. Table Width: 2.5} \\
        \texttt{Num. of Queries: 279}}} & Contriever & 22.8 & 0.7 & 26.1 & 1.1 & 32.0 & 1.8 & 41.4 & 5.4 & 13 \\
      \cmidrule(l){2-11}
      & JAR (MIP) & \gcolor{\textbf{+7.9}} & \gcolor{\textbf{+3.6}} & \gcolor{\textbf{+12.2}} & \gcolor{\textbf{+6.1}} & \gcolor{\underline{+16.4}} & \gcolor{\underline{+8.2}} & \gcolor{\underline{+17.2}} & \gcolor{\underline{+8.6}} & + (45 / 55 / 2206 / 11340) \\
      & \textbf{JAR}$_{\text{iterative}}$ \textbf{(Ours)} & \gcolor{\underline{+6.4}} & \gcolor{\textbf{+3.6}} & \gcolor{\underline{+10.7}} & \gcolor{\underline{+5.7}} & \gcolor{\textbf{+19.8}} & \gcolor{\textbf{+9.0}} & \gcolor{\textbf{+21.2}} & \gcolor{\textbf{+9.3}} & + \textasciitilde10 \\[5pt]
      \midrule

      \multirow{3}{*}{\parbox{2.8cm}{\raggedright \textbf{BEAVER-DW} \\[3pt] \tiny \linespread{0.9}\selectfont
        \texttt{Num. of DBs: 1} \\
        \texttt{Num. of Tables: 97} \\
        \texttt{Avg. Table Width: 15.8} \\
        \texttt{Num. of Queries: 120}}} & Contriever & \textbf{29.3} & \underline{1.7} & \underline{37.5} & \underline{9.2} & \underline{48.5} & \underline{17.5} & \underline{63.5} & \underline{30.8} & 13 \\
      \cmidrule(l){2-11}
      & JAR (MIP) & \rcolor{-21.1} & \rcolor{-1.7} & \rcolor{-13.3} & \rcolor{-0.9} & \textdagger & \textdagger & \textdagger & \textdagger & + (18659 / 34756 / \textdagger / \textdagger) \\
      & \textbf{JAR}$_{\text{iterative}}$ \textbf{(Ours)} & \rcolor{\underline{-0.3}} & \gcolor{\textbf{+0.8}} & \gcolor{\textbf{+4.2}} & \gcolor{\textbf{+3.3}} & \gcolor{\textbf{+4.1}} & \gcolor{\textbf{+6.7}} & \gcolor{\textbf{+1.5}} & \gcolor{\textbf{+2.5}} & + \textasciitilde82 \\[5pt]
      \midrule

      \multirow{3}{*}{\parbox{2.8cm}{\raggedright \textbf{BEAVER-NW} \\[2pt] \tiny \linespread{0.9}\selectfont
        \texttt{Num. of DBs: 5} \\
        \texttt{Num. of Tables: 366} \\
        \texttt{Avg. Table Width: 7.4} \\
        \texttt{Num. of Queries: 86}}} & Contriever & 23.6 & 1.2 & 30.2 & 1.2 & 38.5 & \underline{2.3} & \underline{48.4} & \underline{9.3} & 19 \\ [2pt]
      \cmidrule(l){2-11}
      & JAR (MIP) & \gcolor{\textbf{+10.3}} & \gcolor{0.0} & \gcolor{\textbf{+9.4}} & \gcolor{\underline{+3.5}} & \gcolor{\underline{+10.7}} & \rcolor{-2.3} & \rcolor{-0.4} & \rcolor{-9.3} & + (210 / 572 / 7297 / 10449) \\
      & \textbf{JAR}$_{\text{iterative}}$ \textbf{(Ours)} & \gcolor{\underline{+3.0}} & \gcolor{0.0} & \gcolor{\underline{+6.5}} & \gcolor{\textbf{+4.6}} & \gcolor{\textbf{+11.1}} & \gcolor{\textbf{+8.2}} & \gcolor{\textbf{+6.1}} & \gcolor{\textbf{+3.5}} & + \textasciitilde30 \\[3pt]
      \bottomrule
   \end{tabular}
  \end{adjustbox}
\end{table}

\paragraph{Analysis of Performance and Scalability.}
As shown in Table \ref{tab:retrieval-performance}, our greedy method achieves highly competitive, and often superior, retrieval performance compared to the JAR$_{MIP}$ baseline, while being dramatically faster. For example, on BIRD at K=3, our method achieves 99\% of JAR's CR score in only 0.7\% of the time (101s vs 13679s, a >135x speedup). This efficiency advantage becomes an enabling factor on the complex BEAVER benchmarks. Our iterative method was able to complete all BEAVER-DW runs, while JAR (MIP) timed out on K$\ge$5. On BEAVER-NW (K=5), our method was >240x faster (30s vs 7297s) while achieving a +10.5 point higher CR.

It is also worth noting that even when JAR (MIP) finds a 100\% optimal solution according to its objective function (e.g., on SPIDER at K=2, see Appendix \ref{sec:mip_solver_status}), our iterative method can still achieve slightly higher performance (see Table \ref{tab:retrieval-performance-absolute}). This suggests the MIP's objective is not a perfect proxy for the end retrieval metrics, further motivating the exploration of alternative heuristic formulations.

CRUSH's lower performance stems from two core design differences: it is not explicitly join-aware, and it employs an aggressive coverage-first greedy logic, unlike our holistic gain-based approach. As analyzed in Appendix \ref{sec:crush_failure}, this heuristic is sensitive to noise and can fail to select optimal tables. Given the performance drop on SPIDER and BIRD, we skipped evaluating CRUSH on FIBEN and BEAVER benchmarks.

Finally, the slight performance drop of our method on SPIDER at K=5 and K=10 (relative to the Contriever baseline) can be attributed to the fixed candidate set. As K increases, the re-ranker is forced to include lower-ranked tables from the initial top-20 set, which are more likely to be irrelevant and can create confusion for the selection algorithm on certain queries.

\section{Limitations and Future Work}

This paper proposed reframing multi-table retrieval as an \textbf{iterative, explorative search process}, an alternative to one-shot optimization with inherent advantages in scalability, interpretability, and extensibility. Our \textbf{Greedy Iterative Join-Aware Multi-Table Retrieval algorithm} demonstrated this viability, achieving strong empirical results on standard and complex enterprise benchmarks at a fraction of the computational cost of the MIP-based JAR method \citep{chen-etal-2024-table}.While effective, our greedy approach is sensitive to its initial seed selection; a poor start can lead to a suboptimal final set. 

A crucial direction for future work is to enhance the robustness of the iterative search. The framework's step-by-step nature lends itself well to incorporating \textbf{backtracking mechanisms}. Exploring strategies where the algorithm can revisit earlier decisions if subsequent steps yield low utility could significantly mitigate the risks associated with greedy choices.Furthermore, the framework's true potential lies in its \textbf{extensibility beyond joins}, such as incorporating the UNION operator. This involves modifying the selection utility $U(T_i, \mathcal{C}_k)$ to reward schema compatibility and row diversity. The iterative context $\mathcal{C}_k$ allows for dynamic strategies, like prioritizing joins then exploring unions. Exploring such adaptive strategies, potentially learning operator priority, is a rich research avenue.

Future work must also investigate the framework's robustness to the quality of the initial LLM-based query decomposition and test on more diverse, open-domain datasets with even more complex join paths. Hybrid models, using a fast greedy search by default but triggering a complex solver for specific patterns, also warrant investigation. To provide a broader comparative study, we also plan to evaluate our iterative framework against other recent LLM-Based and schema-routing approaches, such as MURRE~\citep{zhang-etal-2025-murre} and DBCopilot~\citep{DBLP:conf/edbt/WangCLH0WZ25}. We advocate for continued research into practical, scalable, and flexible architectures for multi-table retrieval.

\bibliographystyle{plainnat}
\bibliography{references}

\newpage

\appendix

\section{Ablation Study on Coefficients}
\label{sec:ablation_appendix}
Table \ref{tab:ablation_appendix} details the ablation study on the coefficients for the utility function (Eq. 3) of our iterative greedy algorithm. We tested isolating each component ($\lambda=1$ for one, $0$ for others) and removing each component ($\lambda=1$ for two, $0$ for one). These results informed our Custom Configuration ($\lambda_{\text{coarse}}=4, \lambda_{\text{cov}}=2, \lambda_{\text{join}}=1$), which consistently performs well. The study confirms that coarse relevance is the strongest signal, followed by coverage, and then join compatibility.

\begin{table}[H]
\caption{Ablation study on utility function coefficients. All runs use the top-20 candidates from Contriever, except for FIBEN which uses top-30. Metrics are Recall (R) and Complete Recall (CR) at K. The \textbf{bold} values indicate the best score for each metric within each benchmark, and \underline{underlined} values indicate the second-best.}
\label{tab:ablation_appendix}
\centering
\begin{adjustbox}{width=\textwidth}
\small
\setlength{\tabcolsep}{4pt}
\begin{tabular}{ll rrrr rrrr}
    \toprule
    \multirow{2}{*}{\textbf{Benchmark}} & \multirow{2}{*}{\textbf{Setting ($\lambda_{\text{cov}}, \lambda_{\text{join}}, \lambda_{\text{coarse}}$)}} & \multicolumn{2}{c}{\textbf{R@2 / CR@2}} & \multicolumn{2}{c}{\textbf{R@3 / CR@3}} & \multicolumn{2}{c}{\textbf{R@5 / CR@5}} & \multicolumn{2}{c}{\textbf{R@10 / CR@10}} \\
    \cmidrule(lr){3-4} \cmidrule(lr){5-6} \cmidrule(lr){7-8} \cmidrule(lr){9-10}
    & & R & CR & R & CR & R & CR & R & CR \\
    \midrule
    
    \multirow{7}{*}{SPIDER (k=20)}
    & Only Coverage (1,0,0) & 74.2 & 50.3 & 78.6 & 57.7 & 78.8 & 57.7 & 78.9 & 58.0 \\
    & Only Join (0,1,0) & 71.5 & 41.8 & 84.1 & 67.8 & 86.1 & 72.3 & 89.6 & 79.1 \\
    & Only Coarse (0,0,1) & 81.3 & 59.9 & 93.8 & 85.6 & \underline{98.9} & \underline{97.6} & \textbf{99.7} & \textbf{99.3} \\
    \cmidrule(lr){2-10}
    & No Coarse (1,1,0) & 71.6 & 49.2 & 78.7 & 62.3 & 82.3 & 69.5 & 85.9 & 75.8 \\
    & No Join (1,0,1) & \underline{83.0} & \underline{64.3} & \underline{94.8} & \underline{87.1} & \textbf{99.0} & \textbf{97.8} & \textbf{99.7} & \textbf{99.3} \\
    & No Coverage (0,1,1) & 74.8 & 47.7 & 85.8 & 71.0 & 87.2 & 74.5 & 90.0 & 80.2 \\
    \cmidrule(lr){2-10}
    & \textbf{Custom Config (2,1,4)} & \textbf{85.5} & \textbf{68.0} & \textbf{95.6} & \textbf{90.0} & 98.0 & 95.6 & \underline{98.3} & \underline{96.5} \\
    \midrule

    \multirow{7}{*}{BIRD (k=20)}
    & Only Coverage (1,0,0) & 62.5 & 32.2 & 67.5 & 39.9 & 68.8 & 41.8 & 69.4 & 42.6 \\
    & Only Join (0,1,0) & 56.8 & 26.4 & 66.6 & 42.1 & 79.4 & 62.8 & 90.2 & 81.7 \\
    & Only Coarse (0,0,1) & 63.5 & 34.8 & 76.2 & 56.2 & 85.2 & 71.3 & \underline{95.1} & \underline{89.9} \\
    \cmidrule(lr){2-10}
    & No Coarse (1,1,0) & 66.1 & 39.3 & 75.7 & 55.5 & 84.2 & 71.2 & 92.7 & 86.7 \\
    & No Join (1,0,1) & \underline{72.2} & \underline{47.5} & \underline{84.2} & \underline{69.5} & \textbf{91.1} & \textbf{82.1} & \textbf{96.2} & \textbf{92.6} \\
    & No Coverage (0,1,1) & 59.9 & 31.7 & 73.4 & 56.3 & 83.4 & \underline{71.3} & 93.9 & 89.6 \\
    \cmidrule(lr){2-10}
    & \textbf{Custom Config (2,1,4)} & \textbf{74.2} & \textbf{51.5} & \textbf{84.7} & \textbf{70.5} & \underline{91.0} & \textbf{82.1} & \textbf{96.2} & \textbf{92.6} \\
    \midrule

    \multirow{7}{*}{FIBEN (k=30)}
    & Only Coverage (1,0,0) & 13.6 & 0.7 & 16.2 & 1.4 & 18.1 & 1.4 & 20.5 & 1.4 \\
    & Only Join (0,1,0) & 24.0 & 1.4 & \underline{33.1} & \underline{4.3} & \textbf{53.0} & \textbf{11.8} & 62.0 & 13.6 \\
    & Only Coarse (0,0,1) & 22.8 & 0.7 & 26.1 & 1.1 & 32.0 & 1.8 & 41.4 & 5.4 \\
    \cmidrule(lr){2-10}
    & No Coarse (1,1,0) & 17.4 & \underline{2.5} & 29.6 & 7.2 & 48.8 & 11.1 & \textbf{62.6} & \textbf{14.7} \\
    & No Join (1,0,1) & 23.7 & 1.1 & 28.9 & 2.5 & 34.6 & 5.0 & 44.9 & 7.9 \\
    & No Coverage (0,1,1) & \underline{26.2} & 1.8 & 32.8 & \underline{4.3} & 51.6 & 9.0 & 62.0 & 13.6 \\
    \cmidrule(lr){2-10}
    & \textbf{Custom Config (2,1,4)} & \textbf{29.2} & \textbf{4.3} & \textbf{36.8} & \textbf{6.8} & \underline{51.8} & \underline{10.8} & \textbf{62.6} & \textbf{14.7} \\
    \midrule
    
    \multirow{7}{*}{BEAVER-DW (k=20)}
    & Only Coverage (1,0,0) & 24.4 & \textbf{2.5} & 31.5 & 4.2 & 37.7 & 7.5 & 42.8 & 8.3 \\
    & Only Join (0,1,0) & 25.6 & \textbf{2.5} & 37.7 & 7.5 & 44.4 & 13.3 & 56.7 & 22.5 \\
    & Only Coarse (0,0,1) & \textbf{29.3} & \underline{1.7} & 37.5 & 9.2 & 48.5 & 17.5 & 63.5 & 30.8 \\
    \cmidrule(lr){2-10}
    & No Coarse (1,1,0) & 25.6 & \textbf{2.5} & 38.0 & \underline{11.7} & 46.4 & \underline{20.0} & 61.0 & 29.2 \\
    & No Join (1,0,1) & 26.0 & \textbf{2.5} & 37.4 & 9.2 & \underline{50.8} & \underline{20.0} & \textbf{66.3} & \textbf{34.2} \\
    & No Coverage (0,1,1) & 27.4 & \textbf{2.5} & \underline{39.8} & 9.2 & 48.4 & 18.3 & 59.2 & 26.7 \\
    \cmidrule(lr){2-10}
    & \textbf{Custom Config (2,1,4)} & \underline{29.0} & \textbf{2.5} & \textbf{41.7} & \textbf{12.5} & \textbf{52.6} & \textbf{24.2} & \underline{65.0} & \underline{33.3} \\
    \midrule

    \multirow{7}{*}{BEAVER-NW (k=20)}
    & Only Coverage (1,0,0) & \underline{27.9} & 0.0 & 32.3 & 1.2 & 36.4 & 1.2 & 40.8 & 5.8 \\
    & Only Join (0,1,0) & 27.3 & \textbf{2.3} & 33.8 & \textbf{5.8} & 41.2 & 8.1 & 50.8 & \underline{11.6} \\
    & Only Coarse (0,0,1) & 23.6 & \underline{1.2} & 30.2 & 1.2 & 38.5 & 2.3 & 48.4 & 9.3 \\
    \cmidrule(lr){2-10}
    & No Coarse (1,1,0) & 26.2 & \textbf{2.3} & 34.6 & \underline{4.7} & \underline{45.6} & 7.0 & \underline{53.7} & 10.5 \\
    & No Join (1,0,1) & \textbf{28.2} & 0.0 & \underline{36.4} & 1.2 & 44.1 & 2.3 & 50.4 & 8.1 \\
    & No Coverage (0,1,1) & 26.7 & \underline{1.2} & 33.3 & \textbf{5.8} & 40.7 & \underline{8.1} & 50.2 & \underline{11.6} \\
    \cmidrule(lr){2-10}
    & \textbf{Custom Config (2,1,4)} & 26.6 & \underline{1.2} & \textbf{36.7} & \textbf{5.8} & \textbf{49.6} & \textbf{10.5} & \textbf{54.5} & \textbf{12.8} \\
    \bottomrule
\end{tabular}
\end{adjustbox}
\end{table}

\section{MIP Solver Status}
\label{sec:mip_solver_status}

Table \ref{tab:solver_status_appendix} provides a detailed breakdown of the JAR (MIP) solver's termination status. The experiments were run on a server equipped with Intel(R) Xeon(R) Gold 6330 CPUs @ 2.00GHz, using an allocated resource quota of 4 CPU cores (8 threads) and 64 GB of RAM for this task. We used the Python-MIP library, similar to \cite{chen-etal-2024-table}, and the statuses correspond to its official numeric codes\footnote{\url{https://python-mip.readthedocs.io/en/latest/classes.html}}. The data shows that while most simple queries (SPIDER K=2, BIRD K=2) were solved to optimality, timeouts become common as complexity increases. This results in either a suboptimal \texttt{Feasible} solution (one was found, but not proven optimal) or \texttt{No Solution} (no solution was found within the 1-hour limit). Timeouts were prevalent on BIRD (K=5), FIBEN (K=10), BEAVER-NW (K=5, K=10), and most significantly on BEAVER-DW (K=2), which saw a high percentage of timeouts before any solution could be found.

\begin{table}[H]
  \caption{JAR (MIP) solver status breakdown for multi-table queries (3600s timeout/query). Percentages relative to total multi-table queries (SPIDER: 459, BIRD: 1172, FIBEN: 279, BEAVER-NW: 86, BEAVER-DW: 120).}
  \label{tab:solver_status_appendix}
  \centering
  \small
  \setlength{\tabcolsep}{5pt}
  \begin{tabular}{ll rrrr}
    \toprule
    \textbf{Benchmark} & \textbf{K} & \textbf{Optimal (\%)} & \textbf{Feasible (\%)} & \textbf{No Solution (\%)} & \textbf{Infeasible (\%)} \\
    \midrule
    \multirow{3}{*}{SPIDER} & K=2 & 459 (100.0\%) & 0 (0.0\%) & 0 (0.0\%) & 0 (0.0\%) \\
    & K=3 & 459 (100.0\%) & 0 (0.0\%) & 0 (0.0\%) & 0 (0.0\%) \\
    & K=5 & 459 (100.0\%) & 0 (0.0\%) & 0 (0.0\%) & 0 (0.0\%) \\
    \midrule
    \multirow{3}{*}{BIRD} & K=2 & 1172 (100.0\%) & 0 (0.0\%) & 0 (0.0\%) & 0 (0.0\%) \\
    & K=3 & 1172 (100.0\%) & 0 (0.0\%) & 0 (0.0\%) & 0 (0.0\%) \\
    & K=5 & 1144 (97.6\%) & 24 (2.0\%) & 4 (0.3\%) & 0 (0.0\%) \\
    \midrule
    \multirow{4}{*}{FIBEN} & K=2 & 279 (100.0\%) & 0 (0.0\%) & 0 (0.0\%) & 0 (0.0\%) \\
    & K=3 & 279 (100.0\%) & 0 (0.0\%) & 0 (0.0\%) & 0 (0.0\%) \\
    & K=5 & 279 (100.0\%) & 0 (0.0\%) & 0 (0.0\%) & 0 (0.0\%) \\
    & K=10 & 240 (86.0\%) & 39 (14.0\%) & 0 (0.0\%) & 0 (0.0\%) \\
    \midrule
    \multirow{4}{*}{BEAVER-NW} & K=2 & 86 (100.0\%) & 0 (0.0\%) & 0 (0.0\%) & 0 (0.0\%) \\
    & K=3 & 86 (100.0\%) & 0 (0.0\%) & 0 (0.0\%) & 0 (0.0\%) \\
    & K=5 & 84 (97.7\%) & 2 (2.3\%) & 0 (0.0\%) & 0 (0.0\%) \\
    & K=10 & 50 (58.1\%) & 28 (32.6\%) & 8 (9.3\%) & 0 (0.0\%) \\
    \midrule
    \multirow{1}{*}{BEAVER-DW} & K=2 & 32 (26.7\%) & 4 (3.3\%) & 84 (70.0\%) & 0 (0.0\%) \\
    \multirow{1}{*}{BEAVER-DW} & K=3 & 45 (37.5\%) & 28 (23.3\%) & 47 (39.2\%) & 0 (0.0\%) \\
    \bottomrule
  \end{tabular}
\end{table}

The status codes reported in the table are defined as follows:
\begin{itemize}
    \item \textbf{Optimal:} The MIP solver found and proved the globally optimal solution.
    \item \textbf{Feasible:} The solver found a solution but ran out of time before proving optimality.
    \item \textbf{No Solution Found:} The solver exhausted its time limit before finding any solution.
    \item \textbf{Infeasible:} The constraints are contradictory; no solution can satisfy all of them.
\end{itemize}

\section{Distribution of Multi-Table Queries}
\label{sec:distribution_of_queries}

The following table details the distribution of multi-table queries within each benchmark, categorized by the number of tables required to answer the query. These values were calculated based on the list of tables that were mentioned in the corresponding gold SQL queries. For example, in BIRD, one query requires 8 tables to be answered, so it is counted under the "8" column.

\begin{table}[H]
  \caption{Distribution of multi-table queries based on the number of tables required by the gold SQL query.}
  \label{tab:query-distribution}
  \centering
  \setlength{\tabcolsep}{6pt}
  \begin{tabular}{l rrrrr r}
    \toprule
    \textbf{\# Tables in} & \multicolumn{5}{c}{\textbf{Benchmark}} & \multirow{2}{*}{\textbf{Total}} \\
    \cmidrule(lr){2-6}
    \textbf{Gold Query} & \textbf{SPIDER} & \textbf{BIRD} & \textbf{FIBEN} & \textbf{BEAVER-DW} & \textbf{BEAVER-NW} & \\
    \midrule
    \textbf{2} & 393 & 936 & 77 & 5 & 13 & 1424 \\
    \textbf{3} & 60 & 197 & 26 & 44 & 18 & 345 \\
    \textbf{4} & 6 & 34 & 98 & 37 & 24 & 199 \\
    \textbf{5} & 0 & 1 & 63 & 23 & 4 & 91 \\
    \textbf{6} & 0 & 1 & 5 & 8 & 4 & 18 \\
    \textbf{7} & 0 & 2 & 2 & 3 & 4 & 11 \\
    \textbf{8} & 0 & 1 & 6 & 0 & 1 & 8 \\
    \textbf{9} & 0 & 0 & 0 & 0 & 2 & 2 \\
    \textbf{10} & 0 & 0 & 0 & 0 & 7 & 7 \\
    \textbf{11} & 0 & 0 & 2 & 0 & 1 & 3 \\
    \textbf{12} & 0 & 0 & 0 & 0 & 8 & 8 \\
    \midrule
    \textbf{Total} & \textbf{459} & \textbf{1172} & \textbf{279} & \textbf{120} & \textbf{86} & \textbf{2116} \\
    \bottomrule
  \end{tabular}
\end{table}

\section{Case Study: Analysis of CRUSH Re-ranking Failure}
\label{sec:crush_failure}
To illustrate the performance dip of CRUSH mentioned in Section 4, we analyze a query from the SPIDER benchmark. This case study highlights how CRUSH's greedy, segment-coverage-first logic can prioritize covering diverse segments with irrelevant tables over selecting a set of relevant, coherent tables.

For the query regarding high schooler friendships, the baseline Contriever successfully retrieves the two correct tables in the top-2. CRUSH, however, selects one correct table and one incorrect table, failing the Complete Recall (CR) metric. Table \ref{tab:crush-fail-example} summarizes the results.

\begin{table}[H]
\caption{Retrieval results for the example query. The baseline (Contriever) succeeds, while CRUSH fails.}
\label{tab:crush-fail-example}
\centering
\small
\begin{tabular}{@{}ll@{}}
\toprule
\textbf{Item} & \textbf{Tables} \\
\midrule
Gold Tables & \texttt{network\_1.highschooler}, \texttt{network\_1.friend} \\
\midrule
Baseline Top-2 & \gcolor{\texttt{network\_1.friend} (0.6121), \texttt{network\_1.highschooler} (0.5861)} \quad [\textbf{CR@2: True}] \\
CRUSH Top-2 & \gcolor{\texttt{network\_1.friend} (0.4032)}, \rcolor{\texttt{network\_2.personfriend} (0.3899)} \quad [\textbf{CR@2: False}] \\
\bottomrule
\end{tabular}
\end{table}

The query was decomposed into four segments. The similarity breakdown for the relevant tables against these segments is shown in Table \ref{tab:crush-fail-sim}. Note that the baseline's scores (e.g., 0.5861) are the maximum similarity from this table, which CRUSH uses as input.

\begin{table}[H]
\caption{Similarity breakdown for candidate tables against query segments. Segments are: \textbf{(S1)} \texttt{high\_schooler.name}, \textbf{(S2)} \texttt{high\_schooler.student\_id}, \textbf{(S3)} \texttt{friendship.student\_id}, \textbf{(S4)} \texttt{friendship.friend\_id}.}
\label{tab:crush-fail-sim}
\centering
\begin{adjustbox}{width=\textwidth}
\small
\begin{tabular}{@{}lrrrr@{}}
\toprule
\textbf{Table Name} & \textbf{(S1)} \texttt{...name} & \textbf{(S2)} \texttt{...student\_id} & \textbf{(S3)} \texttt{...student\_id} & \textbf{(S4)} \texttt{...friend\_id} \\
\midrule
\gcolor{\texttt{network\_1.highschooler}} & 0.5313 & \textbf{0.5861} & 0.5490 & 0.5344 \\
\gcolor{\texttt{network\_1.friend}} & 0.5297 & \textbf{0.6121} & 0.6065 & 0.5844 \\
\rcolor{\texttt{network\_2.personfriend}} & 0.5291 & 0.5377 & 0.5500 & 0.5597 \\
\bottomrule
\end{tabular}
\end{adjustbox}
\end{table}

\paragraph{Analysis of Greedy Selection}
The failure is a direct result of CRUSH's greedy, segment-coverage algorithm, which does not re-evaluate aggregate relevance but selects tables one-by-one to "check off" segments.

\begin{itemize}[nosep]
    \item \textbf{Iteration 1: Selecting the First Table} \\
    The algorithm looks for the \textit{single best (table, segment) pair} across all candidate tables and all 4 segments.
    \begin{itemize}[nosep]
        \item \texttt{network\_1.highschooler}'s best score is 0.5861 (on S2).
        \item \texttt{network\_1.friend}'s best score is \textbf{0.6121} (on S2).
        \item \texttt{network\_2.personfriend}'s best score is 0.5597 (on S4).
    \end{itemize}
    \textbf{Decision 1:} The highest score overall is 0.6121. The table \texttt{network\_1.friend} is selected, and segment S2 (\texttt{high\_schooler.student\_id}) is marked as "covered".

    \item \textbf{Iteration 2: Selecting the Second Table} \\
    The algorithm now seeks to cover one of the 3 \textit{remaining} segments (S1, S3, S4). It ignores segment S2 and any table already selected (\texttt{network\_1.friend}). It compares the best scores from \textit{unselected} tables on \textit{uncovered} segments:
    \begin{itemize}[nosep]
        \item For \gcolor{\texttt{network\_1.highschooler}} (Gold): Its best remaining score is 0.5490 (on S3).
        \item For \rcolor{\texttt{network\_2.personfriend}} (Wrong): Its best remaining score is \textbf{0.5597} (on S4).
    \end{itemize}
    \textbf{Decision 2:} Because 0.5597 > 0.5490, CRUSH greedily selects the incorrect table \texttt{network\_2.personfriend} to cover segment S4.
\end{itemize}

\paragraph{Conclusion}
The baseline's \#2 table, \texttt{network\_1.highschooler}, was dropped. Its highest relevance score (0.5861 on S2) was "used up" by the first table selected, which had an even higher score (0.6121) on that same segment. CRUSH's logic then forced it to pick \texttt{network\_2.personfriend} because its score on an \textit{uncovered} segment was marginally higher than \texttt{network\_1.highschooler}'s. This case study exemplifies the risk of a purely coverage-based heuristic, which, as noted in Section 4, can be sensitive to noise and lacks the robust, holistic scoring of our iterative, join-aware approach.

\section{Absolute Performance Scores}

Table \ref{tab:retrieval-performance-absolute} presents the absolute retrieval performance scores (Recall and Complete Recall) for all methods, corresponding to the relative delta values shown in Table \ref{tab:retrieval-performance} in the main paper. For each benchmark and metric, the highest value is in \textbf{bold} and the second-highest is \underline{underlined}.

\begin{table}[htbp]
    \caption{Absolute retrieval performance (R/CR \%) on multi-table queries. K is the number of tables retrieved. All re-ranking methods (JAR, CRUSH, Ours) operate on the same top-20 Contriever candidate set. \textdagger~indicates prohibitive runtime or timeout. \textbf{Highest} and \underline{Second-Highest} scores are marked per column within each benchmark.}
  \label{tab:retrieval-performance-absolute}
  \centering
  \begin{adjustbox}{width=\textwidth}
    \scriptsize 
    \setlength{\tabcolsep}{4pt} 
    \begin{tabular}{ll rr rr rr rr r}
      \toprule
      \multirow{2}{*}{\textbf{Benchmark}} & \multirow{2}{*}{\textbf{Method}} & \multicolumn{2}{c}{\textbf{K=2}} & \multicolumn{2}{c}{\textbf{K=3}} & \multicolumn{2}{c}{\textbf{K=5}} & \multicolumn{2}{c}{\textbf{K=10}} & \multirow{2}{*}{\textbf{Time (seconds)}} \\
      \cmidrule(lr){3-4} \cmidrule(lr){5-6} \cmidrule(lr){7-8} \cmidrule(lr){9-10}
      & & Recall (R) & CR & Recall (R) & CR & Recall (R) & CR & Recall (R) & CR & \\
      \midrule
      
      \multirow{4}{*}{\parbox{2.8cm}{\raggedright \textbf{SPIDER} \\[3pt] \tiny \linespread{0.9}\selectfont
        \texttt{Num. of DBs: 20} \\
        \texttt{Num. of Tables: 81} \\
        \texttt{Avg. Table Width: 5.4} \\
        \texttt{Num. of Queries: 459}}} 
      & Contriever & 81.3 & 59.9 & 93.8 & 85.6 & \underline{98.9} & \underline{97.6} & \textbf{99.7} & \textbf{99.3} & 15 \\
      \cmidrule(l){2-11}
      & JAR (MIP) & \underline{85.4} & \underline{67.1} & \textbf{96.4} & \textbf{91.3} & \textbf{99.4} & \textbf{98.7} & \textdagger & \textdagger & + (325 / 290 / 6360 / \textdagger) \\
      & \textbf{JAR}$_{\text{iterative}}$ \textbf{(Ours)} & \textbf{85.5} & \textbf{68.0} & \underline{95.6} & \underline{90.0} & 98.0 & 95.6 & \underline{98.3} & \underline{96.5} & + \textasciitilde20 \\
      & CRUSH$_{\text{Contriever}}$ & 63.7 & 34.6 & 72.4 & 47.5 & 77.4 & 55.6 & 77.8 & 56.0 & + \textasciitilde38 \\
      \midrule
      
      % --- BIRD SECTION - Absolute Values ---
      \multirow{4}{*}{\parbox{2.8cm}{\raggedright \textbf{BIRD} \\[3pt] \tiny \linespread{0.9}\selectfont
        \texttt{Num. of DBs: 11} \\
        \texttt{Num. of Tables: 75} \\
        \texttt{Avg. Table Width: 10.6} \\
        \texttt{Num. of Queries: 1172}}} & Contriever & 63.5 & 34.8 & 76.2 & 56.2 & 85.2 & 71.3 & \underline{95.1} & \underline{89.9} & 23 \\
      \cmidrule(l){2-11}
      & JAR (MIP) & \textbf{74.4} & \textbf{51.7} & \textbf{85.6} & \textbf{71.9} & \underline{90.4} & \underline{81.7} & \textdagger & \textdagger & + (8501 / 13679 / 40756 / \textdagger) \\
      & \textbf{JAR}$_{\text{iterative}}$ \textbf{(Ours)} & \underline{74.2} & \underline{51.5} & \underline{84.7} & \underline{70.5} & \textbf{91.0} & \textbf{82.1} & \textbf{96.2} & \textbf{92.6} & + \textasciitilde101 \\
      & CRUSH$_{\text{Contriever}}$ & 50.7 & 22.0 & 62.7 & 38.9 & 72.8 & 54.4 & 74.5 & 57.1 & + \textasciitilde116 \\
      \midrule

      \multirow{3}{*}{\parbox{2.8cm}{\raggedright \textbf{FIBEN} \\[3pt] \tiny \linespread{0.9}\selectfont
        \texttt{Num. of DBs: 1} \\
        \texttt{Num. of Tables: 152} \\
        \texttt{Avg. Table Width: 2.5} \\
        \texttt{Num. of Queries: 279}}} & Contriever & 22.8 & 0.7 & 26.1 & 1.1 & 32.0 & 1.8 & 41.4 & 5.4 & 13 \\
      \cmidrule(l){2-11}
      & JAR (MIP) & \textbf{30.7} & \textbf{4.3} & \textbf{38.3} & \textbf{7.2} & \underline{48.4} & 10.0 & \underline{58.6} & 14.0 & + (45 / 55 / 2206 / 11340) \\
      & \textbf{JAR}$_{\text{iterative}}$ \textbf{(Ours)} & \underline{29.2} & \textbf{4.3} & \underline{36.8} & \underline{6.8} & \textbf{51.8} & \textbf{10.8} & \textbf{62.6} & \textbf{14.7} & + \textasciitilde10 \\[5pt]
      \midrule

      \multirow{3}{*}{\parbox{2.8cm}{\raggedright \textbf{BEAVER-DW} \\[3pt] \tiny \linespread{0.9}\selectfont
        \texttt{Num. of DBs: 1} \\
        \texttt{Num. of Tables: 97} \\
        \texttt{Avg. Table Width: 15.8} \\
        \texttt{Num. of Queries: 120}}} & Contriever & \textbf{29.3} & \underline{1.7} & \underline{37.5} & \underline{9.2} & \underline{48.5} & \underline{17.5} & \underline{63.5} & \underline{30.8} & 13 \\
      \cmidrule(l){2-11}
      & JAR (MIP) & 8.2 & 0.0 & 24.2 & 8.3 & \textdagger & \textdagger & \textdagger & \textdagger & + (18659 / 34756 / \textdagger / \textdagger) \\
      & \textbf{JAR}$_{\text{iterative}}$ \textbf{(Ours)} & \underline{29.0} & \textbf{2.5} & \textbf{41.7} & \textbf{12.5} & \textbf{52.6} & \textbf{24.2} & \textbf{65.0} & \textbf{33.3} & + \textasciitilde82 \\[5pt]
      \midrule

      \multirow{3}{*}{\parbox{2.8cm}{\raggedright \textbf{BEAVER-NW} \\[2pt] \tiny \linespread{0.9}\selectfont
        \texttt{Num. of DBs: 5} \\
        \texttt{Num. of Tables: 366} \\
        \texttt{Avg. Table Width: 7.4} \\
        \texttt{Num. of Queries: 86}}} & Contriever & 23.6 & 1.2 & 30.2 & 1.2 & 38.5 & \underline{2.3} & \underline{48.4} & \underline{9.3} & 19 \\ [2pt]
      \cmidrule(l){2-11}
      & JAR (MIP) & \textbf{33.9} & 1.2 & \textbf{39.6} & \underline{4.7} & \underline{49.2} & 0.0 & 48.0 & 0.0 & + (210 / 572 / 7297 / 10449) \\
      & \textbf{JAR}$_{\text{iterative}}$ \textbf{(Ours)} & \underline{26.6} & 1.2 & \underline{36.7} & \textbf{5.8} & \textbf{49.6} & \textbf{10.5} & \textbf{54.5} & \textbf{12.8} & + \textasciitilde30 \\[3pt]
      \bottomrule
    \end{tabular}
  \end{adjustbox}
\end{table}

\end{document}